\documentclass[reprint,pra,twocolumn,aps,showpacs,floatfix]{revtex4-1}
\usepackage[latin1]{inputenc}
\usepackage{bm}
\usepackage[usenames]{color}
\usepackage{multirow}
\usepackage{amssymb}
\usepackage{amsbsy}
\usepackage{amsmath}
\usepackage{graphicx}
\usepackage{epsfig}
\usepackage{placeins}
\makeatletter

\usepackage[plainpages=false,pdfpagelabels,colorlinks=true,linkcolor=red,urlcolor=blue,citecolor=blue,pdftitle={Title},pdfauthor={},pdfdisplaydoctitle=true,pdfduplex=DuplexFlipLongEdge]{hyperref}

\definecolor{darkgreen}{rgb}{0,0.60,.2}

\begin{document}

\title{Emergent eigenstate solution and emergent Gibbs ensemble for \\ expansion dynamics in optical lattices}

\author{Lev Vidmar}
\author{Wei Xu}
\author{Marcos Rigol}
\affiliation{Department of Physics, The Pennsylvania State University, University Park, Pennsylvania 16802, USA}

\begin{abstract}
Within the emergent eigenstate solution to quantum dynamics [\href{https://journals.aps.org/prx/abstract/10.1103/PhysRevX.7.021012}{Phys. Rev. X {\bf 7}, 021012 (2017)}], one can construct a local operator (an emergent Hamiltonian) of which the time-evolving state is an eigenstate. Here we show that such a solution exists for the expansion dynamics of Tonks-Girardeau gases in optical lattices after turning off power-law (e.g., harmonic or quartic) confining potentials, which are geometric quenches that do not involve the boost operator. For systems that are initially in the ground state and undergo dynamical fermionization during the expansion, we show that they remain in the ground state of the emergent local Hamiltonian at all times. On the other hand, for systems at nonzero initial temperatures, the expansion dynamics can be described constructing a Gibbs ensemble for the emergent local Hamiltonian (an emergent Gibbs ensemble). 
\end{abstract}
\maketitle

\section{Introduction}

Nonequilibrium dynamics in isolated quantum systems generally result in states in which observables can be described using equilibrium statistical mechanics (they ``thermalize'')~\cite{dalessio_kafri_16}. Nevertheless, there are many examples of intriguing outcomes of quantum dynamics. Some are the result of designing controllable dynamical protocols to create states of matter that do not exist in equilibrium, e.g., long-lived nonequilibrium states after short photoexcitation pulses~\cite{fausti_tobey_11, stojchevska14} and Floquet states~\cite{mentik_balzer_15, mendozaarenas_17}. Others are the result of having current-carrying states. Within the latter, recent studies have aimed at clarifying the role of (quasi-)local conserved quantities in transport in integrable systems~\cite{prosen11, mierzejewski_prelovsek_14, ilievski_medenjak_16, castroalvaredo_doyon_16, bertini16, deluca_collura_17, ilievski_denardis_17, medenjak_karrasch_17}, and generating interesting current-carrying states in closed quantum systems using inhomogeneous initial states~\cite{antal99, karevski02, rigol04, rigol_muramatsu_05, rigol05a, gobert05, daley05, hm08, santos11, jesenko_znidaric_11, karrasch_moore_13, eisler13, sabetta_misguich_13, alba14, vasseur15, lancaster16, eisler_maislinger_16, ljubotina_znidaric_17}. Other remarkable dynamical phenomena that have recently attracted much attention are dynamical phase transitions~\cite{heyl_polkovnikov_13, vosk14, heyl_14, andraschko14, heyl_15} and discrete time crystals~\cite{yao17, zhang_hess_17, choi_choi_17}.

Also departing from the traditional thermalization scenario, it was pointed out in Ref.~\cite{vidmar_iyer_17} that, for certain classes of quantum quenches involving pure states, there exist a local operator (an emergent local Hamiltonian) of which the time-evolving state is an eigenstate. Namely, there exists an emergent eigenstate solution to the quantum dynamics. Whenever the time-evolving state is the ground state of the emergent local Hamiltonian, one can understand why ground-state-like correlations, such as those observed in Refs.~\cite{rigol04,vidmar15}, can occur far from equilibrium. On the other hand, the emergent eigenstate solution also offers promising tools to engineer many-body states with ultracold atoms. Related ideas have been recently explored in the context of integrable Floquet dynamics~\cite{gritsev17} and counter-diabatic driving~\cite{sels_polkovnikov_17}.

In the quantum quenches studied in Ref.~\cite{vidmar_iyer_17}, the initial states were eigenstates of Hamiltonians that contained the so-called boost operator (an operator that is used in integrable models to generate conserved quantities). While such Hamiltonians could potentially be engineered in optical lattice experiments, they are not directly relevant to current experiments. Our first goal in this paper is to show that the emergent eigenstate solutions also exist for quenches that do not involve the boost operator. Our second goal is to formalize the numerical observation in Ref.~\cite{xu17} that the emergent local Hamiltonian can be used to understand the dynamics of thermal states after a quench. We justify analytically, and show numerically, the applicability of an emergent Gibbs ensemble (a Gibbs ensemble for the emergent local Hamiltonian) for quenches starting from Gibbs states. 

We study bosons in the Tonks-Girardeau regime (hard-core bosons)~\cite{girardeau60,olshanii98,paredes_widera_04,kinoshita_wenger_04} in one-dimensional (1D) lattices, and consider systems that are initially at zero and nonzero temperatures confined in power-law traps (harmonic traps are the ones relevant to current experiments). The physical phenomenon that will be central to our discussions is the dynamical fermionization of the quasimomentum distribution function during the expansion after suddenly turning off the trap~\cite{rigol_muramatsu_05, minguzzi_gangardt_05, rigol05a, xu17}. In closing, by comparing results for power-law traps with even and odd exponents (for spectra that are bounded and unbounded from below, respectively), we show that different emergent local Hamiltonians can be used to describe dynamics of the same states, and that the target eigenstates can be ground states or highly excited states of such Hamiltonians.

\section{Emergent eigenstate solution}

We first review some key elements of the emergent eigenstate solution to quantum dynamics introduced in Ref.~\cite{vidmar_iyer_17}. We consider a quantum quench from the initial Hamiltonian $\hat H_0$ to the final Hamiltonian $\hat H$, with the initial state $|\psi_0\rangle$ being an eigenstate of $\hat H_0$ ($\hat H_0 |\psi_0 \rangle = \lambda |\psi_0 \rangle$). The time-evolving state $|\psi(t) \rangle = e^{-i \hat H t} |\psi_0 \rangle$ is an eigenstate $\hat{\cal M}(t) |\psi(t) \rangle = 0$ of the operator
\begin{equation} \label{def_Heme}
 \hat{\cal M}(t) \equiv e^{-i \hat H t} \hat H_0 e^{i \hat H t} - \lambda \, ,
\end{equation}
which is in general nonlocal and therefore of no particular interest (we set $\hbar=1$). However, there are physically relevant cases for which $\hat{\cal M}(t)$ is a local operator, i.e., an extensive sum of operators with support on a finite number of lattice sites~\cite{vidmar_iyer_17}. We say that the emergent eigenstate solution to quantum dynamics exists whenever $\hat{\cal M}(t)$ can be replaced by a local operator $\hat{\cal H}(t)$ that we call the emergent local Hamiltonian. Since $\hat{\cal H}(t) |\psi(t) \rangle = 0$, instead of time-evolving the initial state one can solve for the dynamics by finding a single eigenstate of $\hat{\cal H}(t)$. Remarkably, $\hat{\cal H}(t)$ is time independent in the Heisenberg picture [$\langle \psi(t) | \hat{\cal H}(t) | \psi(t) \rangle = 0$]. Hence, $\hat{\cal H}(t)$ is a local conserved quantity even though it does not commute with the physical Hamiltonian that governs the dynamics.

The conditions for $\hat{\cal H}(t)$ to exist become apparent from the expansion of Eq.~(\ref{def_Heme}) in power series 
\begin{equation} \label{def_Heme_sum}
 e^{-i \hat H t} \hat H_0 e^{i \hat H t} = \hat H_0 + \sum_{n=1}^{\infty} \frac{(-i t)^n}{n!} \hat{\cal H}_n \, , 
\end{equation}
where $\hat{\cal H}_n = [\hat H,...[\hat H,[\hat H,\hat H_0]]...]$ is an $n$-th nested commutator of the final Hamiltonian with the initial Hamiltonian. An emergent local Hamiltonian exists if: (i) $\hat {\cal H}_n$ vanishes at some finite $n_0$, or (ii) if the nested commutators close the sum in Eq.~(\ref{def_Heme_sum}).

\section{Emergent Gibbs ensemble}

One can generalize the emergent eigenstate solution of quantum dynamics to initial thermal states. For an initial density matrix $\hat \rho_0 = e^{-\beta \hat H_0}/Z_0$, where $Z_0 = {\rm Tr}\{ e^{-\beta \hat H_0} \}$ and $\beta=T^{-1}$ is the initial inverse temperature (we set $k_{\rm B}=1$), the time-evolving density matrix is
\begin{eqnarray}
 \hat \rho(t) & = & Z_0^{-1} e^{-i \hat H t} e^{-\beta \hat H_0} e^{i \hat H t} \nonumber \\
 & = & Z_0^{-1} \sum_{n=0}^{\infty} \frac{(-\beta)^n}{n!} e^{-i \hat H t} \left( \hat H_0 \right)^{\!n} e^{i \hat H t} \, , \label{def_rho_t}
\end{eqnarray}
where, in the second line, the operator $e^{-\beta \hat H_0}$ was expanded in a power series. Writing $e^{-i \hat H t} (\hat H_0)^n e^{i \hat H t} = (e^{-i \hat H t} \hat H_0 e^{i \hat H t} )^n$ yields
\begin{equation} \label{def_rho_general}
 \hat \rho(t) = Z_0^{-1} \exp\left( -\beta \left[ e^{-i \hat H t} \hat H_0 e^{i \hat H t} \right] \right) \, .
\end{equation}
In Eq.~(\ref{def_rho_general}), one can introduce the operator $\hat{\cal M}(t)\equiv e^{-i \hat H t} \hat H_0 e^{i \hat H t}$ such that the time-evolving density matrix $\hat \rho(t)$ is a Gibbs density matrix of $\hat{\cal M}(t)$. $\hat{\cal M}(t)$ is in general nonlocal and, hence, of no particular use. However, whenever $\hat{\cal M}(t)$ is local [$\hat{\cal M}(t)\equiv\hat{\cal H}(t)$], Eq.~(\ref{def_rho_general}) represents a physically meaningful emergent Gibbs ensemble
\begin{equation} \label{def_rho_eme}
 \hat \Sigma(t) = Z_0^{-1} e^{-\beta \hat{\cal H}(t)} \, .
\end{equation}
Note that the temperature in $\hat \Sigma(t)$ is that of the initial state, only the emergent local Hamiltonian changes with time. The expectation value of any observable $\hat O$ during the dynamics can be computed as $\langle \hat O(t) \rangle = {\rm Tr} \{ \hat \Sigma(t) \hat O\}$.

\section{Dynamical fermionization}

As mentioned before, here we study hard-core bosons in 1D lattices. We consider initial Hamiltonians of the form:
\begin{equation} \label{def_H0b}
 \hat H_{0,\text{HCB}}^{(\alpha)}= - J \sum_{l=-L_0}^{L_0-1} \left(\hat b_{l+1}^\dagger \hat b^{}_l + {\rm H.c.} \right) + \frac{J}{R^\alpha} \sum_{l=-L_0}^{L_0} l^\alpha\, \hat b_{l}^\dagger \hat b_l \, ,
\end{equation}
where $\hat b_{l}^\dagger$ ($\hat b^{}_l$) is the creation (annihilation) operator of a hard-core boson at site $l$, $J$ is the hopping amplitude, and $J/R^\alpha$ and $\alpha$ are the strength and exponent of the power-law trap, respectively. The quantity to be kept constant when taking the thermodynamic limit is the so-called characteristic density $\tilde{\rho}=N/R$~\cite{rigol_muramatsu_04, rigol_muramatsu_05a}, where $N$ is the total number of particles in the trap.

\begin{figure}[!t]
\centering
\includegraphics[width=0.99\columnwidth]{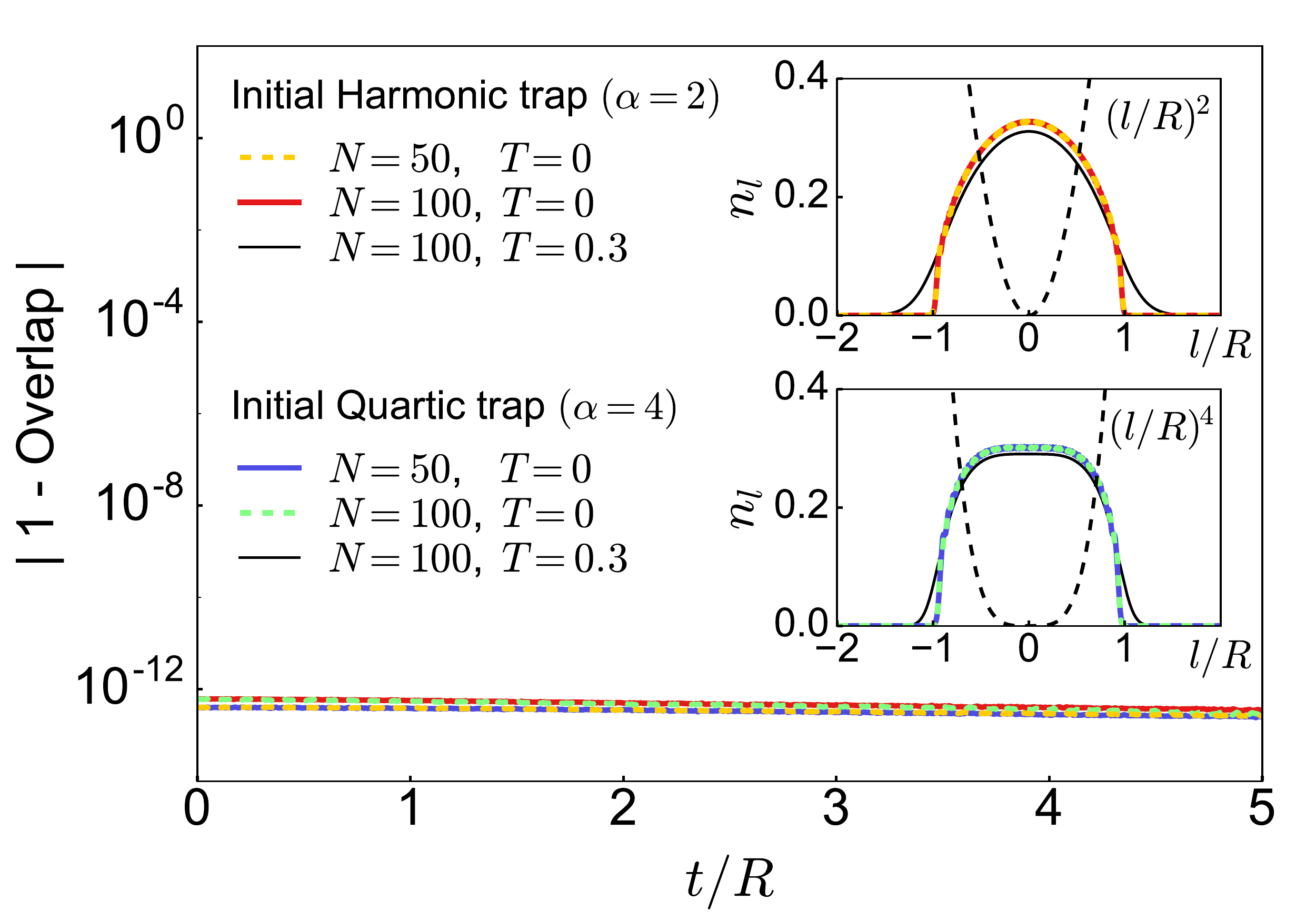}
\caption{
{\it Initial site occupations and validity of the emergent eigenstate solution.}
(Insets) Initial site occupations in harmonic (top inset) and quartic (bottom inset) traps at zero and nonzero temperature. Results are shown for a characteristic density $\tilde{\rho} = N/R = 0.5$. (Main panel) Subtracted overlap $|1-O(t)|$, where $O(t) = |\langle \Psi_t | \psi(t) \rangle|$, of the time-evolving state $|\psi(t)\rangle$ (for the $T=0$ cases) with the ground state $|\Psi_t \rangle$ of the emergent local Hamiltonian $\hat {\cal H}^{(2)}(t)$~(\ref{def_Heme_V2}) and $\hat {\cal H}^{(4)}(t)$~(\ref{def_Heme_V4}).
}\label{fig1}
\end{figure}

Mapping hard-core bosons onto spins 1/2 and spins 1/2 onto fermions, $\hat b_l = e^{i \pi \sum_{m<l} \hat f_m^\dagger \hat f_m} \hat f_l $~\cite{jordan_wigner_28, holstein_primakoff_40, cazalilla_citro_review_11}, Hamiltonian \eqref{def_H0b} maps onto a Hamiltonian of noninteracting spinless fermions
\begin{equation} \label{def_H0}
 \hat H_0^{(\alpha)} = - J \sum_{l=-L_0}^{L_0-1} \left( \hat f_{l+1}^\dagger \hat f_l + {\rm H.c.} \right) + \frac{J}{R^\alpha} \sum_{l=-L_0}^{L_0} l^\alpha \hat f_{l}^\dagger \hat f_l \, .
\end{equation}
One can efficiently compute one-body observables of hard-core bosons solving for the fermions and using properties of Slater determinants~\cite{rigol_muramatsu_05a, rigol05a, rigol05b, xu17}. The site occupations of fermions $n_l = \langle \hat n_l \rangle$, with $\hat n_l=\hat f_{l}^\dagger \hat f_l$, and hard-core bosons are identical.

We first focus on the cases in which $\alpha = 2$ (harmonic trap) and $\alpha = 4$ (quartic trap). The insets in Fig.~\ref{fig1} show typical initial ground-state and finite-temperature ($T = 0.3J$) site occupations considered in our study. It is interesting to note that, in the center of the quartic trap, the site occupations are almost constant.

\begin{figure*}[!tb]
\centering
\includegraphics[width=2\columnwidth]{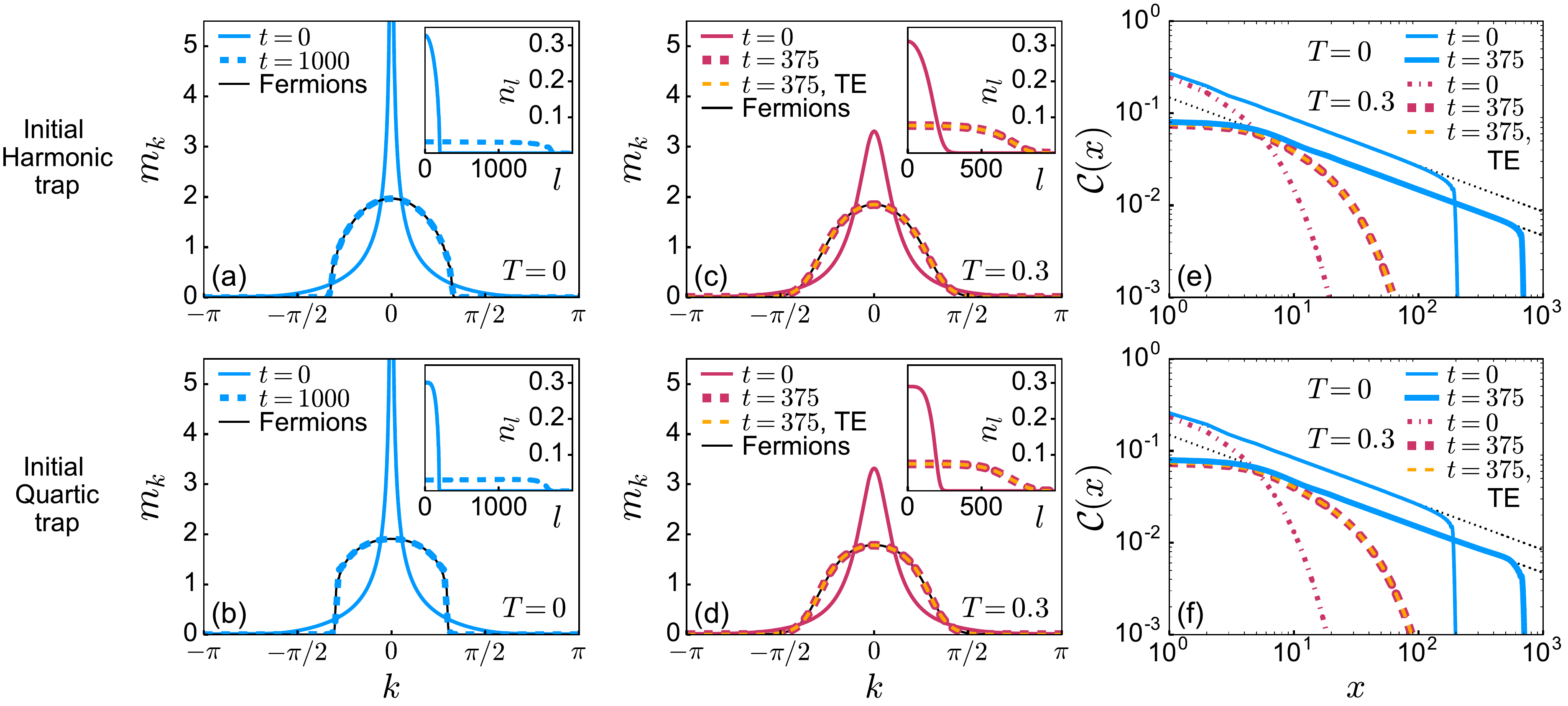}
\caption{
{\it Dynamical fermionization of the hard-core boson momentum distribution function.}
All the results reported are obtained using the emergent local Hamiltonian $\hat {\cal H}^{(2)}(t)$~(\ref{def_Heme_V2}) [upper panels] and $\hat {\cal H}^{(4)}(t)$~(\ref{def_Heme_V4}) [lower panels]. The only exceptions are the thin dashed lines in (c)--(f), which depict results for the time evolution (TE) of thermal ($T=0.3$) initial states. (a)--(d) Quasi-momentum distribution $m_k$ (main panels) and site occupations $n_l$ (insets) in the ground state of the emergent local Hamiltonian [(a),(b)] and in the emergent Gibbs ensemble [(c),(d)]. (e),(f) Absolute value of one-body correlations ${\cal C}(x)=|\langle b_{l=0}^\dagger \hat b_{l'=x}\rangle|$ in the ground state of the emergent local Hamiltonian and in the emergent Gibbs ensemble. Thin dotted lines are power-law fits to ${\cal C}(x) = a x^{-1/2}$ for the ground-state results in the interval $x\in[10,100]$ at $t=0$ and $x \in [50,500]$ at $t=375$. For the harmonic trap, we get $a = 0.270$ at $t=0$ and $a=0.148$ at $t=375$, while for the quartic trap, we get $a = 0.264$ at $t=0$ and $a=0.149$ at $t=375$. The results reported are for systems with $N=100$ particles and a characteristic density $\tilde{\rho} = 0.5$.
}\label{fig2}
\end{figure*}

Our quench consists of turning off the trap, so that the dynamics occur under the physical Hamiltonian
\begin{equation} \label{def_H}
 \hat H = - J \sum_{l=-L_0}^{L_0-1} \left( \hat f_{l+1}^\dagger \hat f^{}_l + {\rm H.c.} \right) \, .
\end{equation}
We measure time in units of $1/J$ and set $J \equiv 1$. This class of geometric quantum quenches is known as sudden expansion and has been widely studied theoretically~\cite{rigol04, rigol_muramatsu_05, rigol05a, rodriguez_manmana_06, delcampo06,delcampo08, buljan_pezer_08, hm08, jukic09, hm09, eisler09, lancaster10, kajala11, bolech12, iyer12, collura_sotiriadis_13, collura13b, vidmar13, boschi14, campbell_gangardt_15, brandino15, hauschild15, ren_wu_15, schluenzen_harmanns_16, mei16, robinson_caux_16, xu17} and in experiments with ultracold atoms in optical lattices~\cite{schneider12, ronzheimer13, xia_zundel_15, vidmar15,preiss_ma_15}. In contrast to the standard time-of-flight measurements, the lattice potential and, as a result, strong interactions, are not switched off during the quench.

We now derive the emergent local Hamiltonian for this setup. Note that, in Ref~\cite{vidmar_iyer_17}, the initial and the final Hamiltonians satisfied the commutation relation $[ \hat H, \hat H_0 ] \propto \hat Q$, with $\hat Q$ being a conserved quantity of the final Hamiltonian (up to boundary terms). Such a commutation relation was enforced by making $\hat H_0$ the sum of $\hat H$ and a boost operator. For $\alpha =2$ and 4 [see Eq.~\eqref{def_H0}], the traps of interest here, this is not the case. 

For the analysis that follows, it is useful to define the generalized kinetic energy $\hat T^{(m,\alpha)}$ and current $\hat J^{(m,\alpha)}$ operators
\begin{eqnarray}
 \hat T^{(m,\alpha)} & = & - R^{-\alpha} \sum_{l=-L_0}^{L_0-m} \left[ \left(l+\frac{m}{2}\right)^\alpha \hat f_{l+m}^\dagger \hat f_{l} + {\rm H.c.} \right], \\
 \hat J^{(m,\alpha)} & = & R^{-\alpha} \sum_{l=-L_0}^{L_0-m} \left[ i \left(l+\frac{m}{2}\right)^\alpha \hat f_{l+m}^\dagger \hat f_{l} - {\rm H.c.} \right].
\end{eqnarray}
They allow us to write the initial Hamiltonian as $\hat H^{(\alpha)}_0 = \hat T^{(1,0)} - \hat T^{(0,\alpha)}/2$, and the final Hamiltonian as $\hat H = \hat T^{(1,0)}$. Note that the operators $\hat T^{(m,0)}$ and $\hat J^{(m,0)}$ commute with $\hat H$, up to boundary terms. This is what ensures that the sum in Eq.~(\ref{def_Heme_sum}) is convergent for our quenches of interest.

Let us consider first the initial harmonic confinement, $\alpha = 2$ in Eq.~(\ref{def_H0}). The commutator of the final and the initial Hamiltonian yields
\begin{equation}
 [\hat H, \hat H_0^{(2)}] = -2i \, R^{-1} \hat J^{(1,1)} \, .
\end{equation}
While $\hat J^{(1,1)}$ does not commute with $\hat H$, their commutator yields the operators
\begin{equation} \label{def_comm_H_J11}
 [\hat H, \hat J^{(1,1)}] = -i \, R^{-1} \hat T^{(2,0)} + 2i \, R^{-1} \hat N \, ,
\end{equation}
with $\hat N=\sum_l\hat n_l$, which both commute with $\hat H$ (up to boundary terms). We therefore truncate the series in Eq.~(\ref{def_Heme_sum}) at $n_0 = 2$. Consequently, the emergent local Hamiltonian for the initial harmonic trap reads
\begin{equation} \label{def_Heme_V2}
 \hat{\cal H}^{(2)}(t) = \hat H_0^{(2)} - \lambda - \frac{2 t}{R} \hat J^{(1,1)} + \left( \frac{t}{R} \right)^{\!\!2} \hat T^{(2,0)} + \left( \frac{\sqrt{2}t}{R} \right)^{\!\!2} \!\! \hat N \, .
\end{equation}
Remarkably, when mapping spinless fermions back to hard-core bosons, the operator $\hat T^{(2,0)}$ in Eq.~(\ref{def_Heme_V2}) becomes a two-body operator $-\sum_l (\hat b_{l+2}^\dagger [1-2\hat b_{l+1}^\dagger \hat b_{l+1}] \hat b_l + {\rm H.c.})$. Hence, the emergent local Hamiltonian for hard-core bosons contains correlated next-nearest neighbor hoppings, even though the emergent local Hamiltonian for the corresponding fermions is quadratic.

The emergent local Hamiltonian to describe the expansion dynamics from other initial power-law traps, with $\alpha>2$, can also be determined in a straightforward way. However, the calculations become increasingly lengthy with increasing $\alpha$ as the series in Eq.~(\ref{def_Heme_sum}) is truncated at $n = \alpha$. In Appendix~\ref{app2}, we discuss the calculation for the quartic trap, $\alpha = 4$ in Eq.~(\ref{def_H0}). The resulting emergent local Hamiltonian reads
\begin{widetext}
\begin{eqnarray}
 \hat{\cal H}^{(4)}(t) & = & \hat H_0^{(4)} - \lambda - 6 \left( \frac{t}{R} \right)^{\!\!2} \hat T^{(0,2)}
 + \left( \frac{t}{R} \right)^{\!\!2} \left[ 4 \left( \frac{t}{R} \right)^{\!\!2} + R^{-2} \right] \hat T^{(2,0)} 
 + 6 \left( \frac{t}{R} \right)^{\!\!2} \hat T^{(2,2)} - \left( \frac{t}{R} \right)^{\!\!4} \hat T^{(4,0)} \nonumber \\
 &&
 - \frac{t}{R} \left[ 12 \left( \frac{t}{R} \right)^{\!\!2} + R^{-2} \right] \hat J^{(1,1)} - 4 \left( \frac{t}{R} \right) \hat J^{(1,3)} + 4 \left( \frac{t}{R} \right)^{\!\! 3} \hat J^{(3,1)}
 + 2  \left( \frac{t}{R} \right)^{\!\!2} \left[ 3 \left( \frac{t}{R} \right)^{\!\!2} + R^{-2} \right] \hat N \, , \label{def_Heme_V4}
\end{eqnarray}
\end{widetext}
which is significantly more complex than $\hat{\cal H}^{(2)}(t)$. However, it still consists of sums of local operators.

The emergent local Hamiltonian construction implies that if the initial state is the ground state of $\hat H_0^{(\alpha)}$~(\ref{def_H0}) [for $\alpha$ even and positive] and $\hat{\cal H}^{(\alpha)}(t)$ is nondegenerate, the time-evolving state $|\psi(t)\rangle$ is the ground state of $\hat{\cal H}^{(\alpha)}(t)$ at all times, which we denote as $|\Psi_t\rangle$. This statement is not limited to a particular initial characteristic density $\tilde{\rho}$. We check this numerically for $\alpha=2$ and 4, and $\tilde{\rho} = 0.5$ in Fig.~\ref{fig1} by calculating the overlap $O(t) = |\langle \Psi_t | \psi(t) \rangle |$. The overlap is essentially one to machine precision. This confirms that, if the lattice is sufficiently large such that the site occupations at the edges vanish at all times, the emergent eigenstate solution is valid at all times (see Appendix~\ref{app1}). Moreover, Figs.~\ref{fig2}(c)--\ref{fig2}(f) show results for various one-body observables for initial finite-temperature states, obtained both by time-evolving the initial density matrix (as in Ref.~\cite{xu17}) and by using the emergent Gibbs ensemble. The results can be seen to agree, which demonstrates the validity of the emergent Gibbs ensemble description [see Eq.~(\ref{def_rho_eme})].

Physically, the quench dynamics under investigation is of particular interest because the quasimomentum distribution of hard-core bosons undergoes a dynamical fermionization, namely, it approaches that of spinless fermions as the cloud expands~\cite{rigol_muramatsu_05}. (Note that the quasimomentum distribution of the spinless fermions does not change in time because the fermionic occupations of the quasimomentum modes are conserved quantities.) This dynamical fermionization is to be contrasted to the result of the time-of-flight protocol, in which all the external potentials are switched off and the measured momentum distribution after expansion is, up to higher-order Bragg peaks, the same as the initial quasimomentum distribution of the hard-core bosons. Dynamical fermionization has been studied for hard-core bosons in initial ground states in a lattice~\cite{rigol_muramatsu_05, rigol05a}, in the continuum~\cite{minguzzi_gangardt_05}, and at finite temperatures in a lattice~\cite{xu17}.

Figures~\ref{fig2}(a)--\ref{fig2}(d) display the quasimomentum distribution $m_k$ for $\alpha=2$ and $4$. We define $m_k$ of hard-core bosons as $m_k = (1/R)\sum_{l,l'} e^{ik(l-l')} \langle \hat b_l^\dagger \hat b^{}_{l'} \rangle$ (for spinless fermions, one just needs to replace $\hat b_l^\dagger \hat b^{}_{l'}$ by $\hat f_l^\dagger \hat f^{}_{l'}$). In equilibrium, $m_k$ of hard-core bosons is clearly different from its spinless fermion counterpart: it is sharply peaked at $k=0$ in the ground state [Figs.~\ref{fig2}(a) and~\ref{fig2}(b)], while finite temperatures [Figs.~\ref{fig2}(c) and~\ref{fig2}(d)] smoothen the peak~\cite{rigol05b}. Remarkably, during the dynamics [see Figs.~\ref{fig2}(a)--\ref{fig2}(d)], $m_k$ of hard-core bosons approaches the one of spinless fermions irrespective of the initial temperature~\cite{xu17} and of the exponent of the initial confining potential. The quasimomentum distribution of hard-core bosons is nearly identical to that of the fermions at the longest expansion times studied, when the occupations $n_l$ in the center of the lattice are strongly reduced from their initial values [see the insets of Figs.~\ref{fig2}(a)-\ref{fig2}(d)].

The dynamical fermionization of the bosonic quasimomentum distribution function is not the only intriguing phenomenon that occurs during the expansion dynamics. Another remarkable feature is the preservation of coherence in the many-body wavefunction of pure states. This can be seen by studying the spatial decay of the absolute value of one-body correlations ${\cal C}(x) = | \langle \hat b_{l=0}^\dagger \hat b_{l'=x} \rangle|$, for which results are depicted in Figs.~\ref{fig2}(e) and~\ref{fig2}(f). While the spatial decay for finite-temperature initial states is exponential~\cite{xu17}, it is intriguing that for initial ground states the correlations retain a power-law decay with the ground-state exponent ${\cal C}(x) \propto x^{-1/2}$ at all times \cite{rigol_muramatsu_05, rigol05a}. The fact that the expanding states that start their dynamics from ground states are eigenstates of emergent local Hamiltonians~(\ref{def_Heme_V2}) and~(\ref{def_Heme_V4}) allow one to gain an intuitive understanding for why correlations can be power law. This is a behavior typical of gapless 1D systems in their ground states, which are described by the Luttinger liquid theory~\cite{cazalilla_citro_review_11}.

\section{Nonuniqueness of the emergent local Hamiltonian}

Before concluding, let us also consider initial Hamiltonians of the form
\begin{equation}
\hat H_0'^{(\alpha)} = \sum_{l=-L_0}^{L_0} l^\alpha \hat n_{l} \, ,
\end{equation}
which commute for different values of $\alpha$. Let us consider two initial Hamiltonians $\hat H_0'^{(\alpha_1)}$ and $\hat H_0'^{(\alpha_2)}$ and the same final Hamiltonian $\hat H$ after the quench. The corresponding emergent local Hamiltonians $\hat{\cal H}'^{(\alpha_1)}(t)$ and $\hat{\cal H}'^{(\alpha_2)}(t)$ commute [this follows from Eq.~\eqref{def_Heme}], i.e., they share eigenstates. This means that different emergent local Hamiltonians can be used to describe dynamics from initial states that are common eigenstates of $\hat H_0'^{(\alpha_1)}$ and $\hat H_0'^{(\alpha_2)}$. 

\begin{figure}[!t]
\centering
\includegraphics[width=0.99\columnwidth]{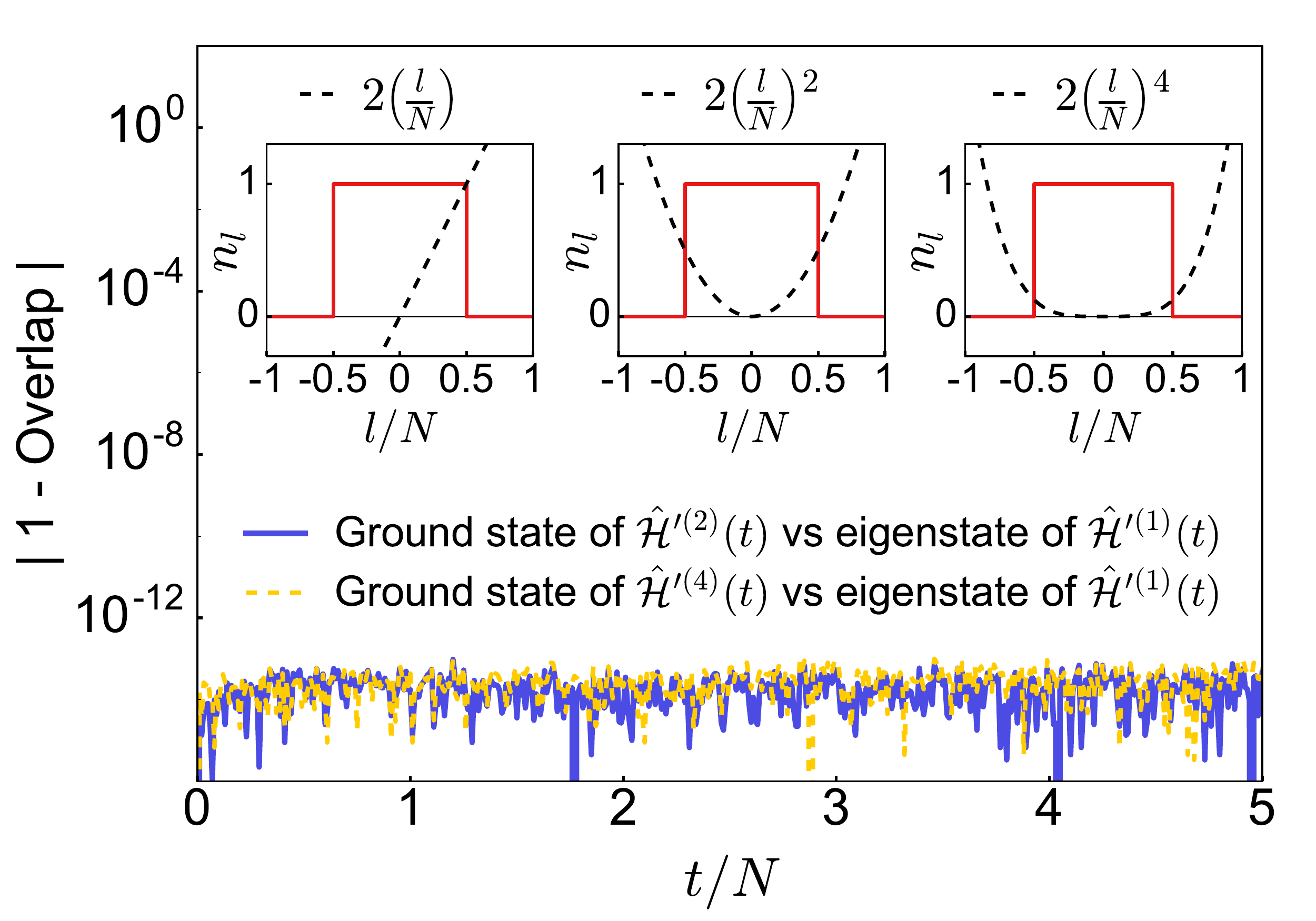}
\caption{
{\it Nonuniqueness of the emergent local Hamiltonian.}
We consider an initial product state $|\psi_{0} \rangle = \prod_{l \in L_{\rm c}} \hat b_l^\dagger |\emptyset\rangle$, where the particles occupy $L_{\rm c}$ consecutive sites in the center of the lattice. This state is an eigenstate of $\hat H_0'^{(1)}$, $\hat H_0'^{(2)}$, and $\hat H_0'^{(4)}$ (see insets for the corresponding site occupations). In the quench, the trap is turned off and hopping between nearest neighbor sites is turned on. As a result, a highly excited eigenstate of the emergent local Hamiltonian $\hat {\cal H}'^{(1)}(t)$ is identical to the ground states of $\hat {\cal H}'^{(2)}(t)$ and $\hat {\cal H}'^{(4)}(t)$, as shown in the main panel by the vanishing values of the subtracted overlap between these states. Results are reported for systems with $N=100$ and $L=2500$.
}\label{fig3}
\end{figure}

An interesting aspect about the nonuniqueness of the emergent local Hamiltonian appears when studying $\hat H_0'^{(\alpha)}$ for even and odd values of $\alpha$, as the former (latter) exhibits a spectrum that is bounded (unbounded) from below. If one considers an initial state that is a Fock state with one particle per site in the center of the lattice, see the insets in Fig.~\ref{fig3}, that state is the ground state of $\hat H_0'^{(2)}$ and $\hat H_0'^{(4)}$, while it is a highly-excited (degenerate) eigenstate of $\hat H_0'^{(1)}$. As a result, the expansion dynamics can be described using the ground state of $\hat {\cal H}'^{(2)}(t)$ and $\hat {\cal H}'^{(4)}(t)$ [obtained by replacing $\hat H_0^{(\alpha)} \rightarrow \hat H_0'^{(\alpha)}$ and $R\rightarrow 1$ in Eqs.~\eqref{def_Heme_V2} and~\eqref{def_Heme_V4}, respectively] or using a highly-excited eigenstate of
\begin{equation}
\hat{\cal H}'^{(1)}(t) = \hat H_0'^{(1)} - t \, \hat J^{(1,0)} - \lambda \, ,
\end{equation}
with $\lambda=0$~\cite{vidmar_iyer_17}. In the main panel in Fig.~\ref{fig3}, we show overlaps between the target eigenstates of $\hat{\cal H}'^{(1)}(t)$, $\hat {\cal H}'^{(2)}(t)$, and $\hat {\cal H}'^{(4)}(t)$ as a function of time. The overlaps are one within machine precision for all times shown.

This example offers an understanding for why one-body correlations with ground-state character can be found in highly-excited eigenstates of emergent local Hamiltonians with spectra that are unbounded from below~\cite{vidmar_iyer_17}. In the case considered here, a highly-excited eigenstate of $\hat{\cal H}'^{(1)}(t)$ is the ground state of $\hat{\cal H}'^{(2)}(t)$ and $\hat{\cal H}'^{(4)}(t)$. Note that, actually, a macroscopic number of low-energy eigenstates of $\hat{\cal H}'^{(2)}(t)$ and $\hat{\cal H}'^{(4)}(t)$ appears (is ``cloned'') in the middle of the spectrum of $\hat{\cal H}'^{(1)}(t)$.

\section{Conclusions}

We have shown that emergent eigenstate solutions exist for the expansion dynamics of Tonks-Girardeau gases in 1D optical lattices after turning off power-law confining potentials. Those quenches do not involve the boost operator, which was central to the discussion in Ref.~\cite{vidmar_iyer_17}. Our construction is applicable independently of the characteristic density chosen, the exponent of the power-law traps, the initial temperature, and for arbitrarily long times, so long as particles do not reach the edges of the lattice. 
The emergent local Hamiltonians constructed here provide a promising tool to manipulate time-evolving states in optical lattices. For example, by quenching to the emergent Hamiltonian during the expansion one can suddenly freeze the atomic cloud, as it becomes a stationary state. This opens a door towards the efficient engineering of tailored many-particle states.

We studied the dynamical fermionization of the hard-core boson quasimomentum distribution function during expansion dynamics, and showed that the dynamically fermionized state can be the ground state of an emergent local Hamiltonian with competing (generalized) kinetic and current operators. We also formally introduced the concept of the emergent Gibbs ensemble to describe quantum dynamics of initial Gibbs states. It can be used to describe the expansion dynamics of Tonks-Girardeau gases, relevant to current experiments with ultracold atomic bosons in optical lattices.

\appendix
\section{Times of validity of the emergent eigenstate description} \label{app1}

In the derivation of $\hat{\cal H}^{(2)}(t)$ and $\hat{\cal H}^{(4)}(t)$, we neglected boundary terms that appear in the commutators $\hat{\cal H}_n$. These terms enter the series in Eq.~(\ref{def_Heme_sum}) generating operators at the lattice boundaries whose support increases with the power of $t$, and eventually result in a breakdown of the emergent eigenstate description of the dynamics on finite lattices and long times. Physically, the breakdown time can be understood to be the time at which propagating particles reach the lattice boundaries~\cite{vidmar_iyer_17}. Hence, by taking limits appropriately, no breakdown of the emergent eigenstate solution will occur. One needs to first take the lattice size to infinity while keeping the time fixed, and then take the infinite time limit.

\section{Emergent local Hamiltonian for the initial quartic trap} \label{app2}

The emergent local Hamiltonian $\hat{\cal H}^{(4)}(t)$ in Eq.~(\ref{def_Heme_V4}) is derived by evaluating elements $\hat{\cal H}_n^{(4)}$ of the series in Eq.~(\ref{def_Heme_sum}). The first-order term is
\begin{equation}
 \hat{\cal H}_1^{(4)} = - i R^{-3} \hat J^{(1,1)} - 4i \, R^{-1} \hat J^{(1,3)} \, ,
\end{equation}
the second-order term is
\begin{equation}
 \hat{\cal H}_2^{(4)} = 24 R^{-2} \hat T^{(0,2)} - 2 R^{-4} \hat T^{(2,0)} - 12 R^{-2} \hat T^{(2,2)} - 4R^{-4}\hat N ,
\end{equation}
the third-order term is
\begin{equation}
 \hat{\cal H}_3^{(4)} = 72i \, R^{-3} \hat J^{(1,1)} - 24 i \, R^{-3} \hat J^{(3,1)} \, ,
\end{equation}
and the fourth-order term is
\begin{equation}
 \hat{\cal H}_4^{(4)} = 96 R^{-4} \hat T^{(2,0)} - 24 R^{-4} \hat T^{(4,0)} + 144 R^{-4} \hat N \, .
\end{equation}
One then realizes that $\hat{\cal H}_4^{(4)}$ commutes with $\hat H$ (up to boundary terms), and, hence, the series in Eq.~(\ref{def_Heme_sum}) can be truncated at $n_0=4$ to produce the emergent local Hamiltonian $\hat{\cal H}^{(4)}(t)$ in Eq.~(\ref{def_Heme_V4}).

\begin{acknowledgements}
We acknowledge discussions with M. Heyl and D. Iyer, and we thank F. Heidrich-Meisner for a careful reading of the manuscript.
This work was supported by the Office of Naval Research, Grant No.~N00014-14-1-0540. The computations were done at the Institute for CyberScience at Penn State.
\end{acknowledgements}

\bibliographystyle{biblev1}
\bibliography{references}

\end{document}